\documentclass[twocolumn,secnumarabic,amssymb, nobibnotes, aps, prd,10pt,superscriptaddress]{revtex4}

\usepackage{graphicx}
\usepackage{float}

\newcommand{\unit}[1]{\ensuremath{\, \mathrm{#1}}}
\begin{document}

\title{Ultrafast charge-transfer exciton dynamics in C$_{60}$ thin films}

\author{Sebastian Emmerich}
\email{emmerich@physik.uni-kl.de}
\affiliation{University of Kaiserslautern and Research Center OPTIMAS, Erwin-Schr\"odinger-Stra\ss{}e 46, 67663 Kaiserslautern, Germany}
\author{Sebastian Hedwig}
\affiliation{University of Kaiserslautern and Research Center OPTIMAS, Erwin-Schr\"odinger-Stra\ss{}e 46, 67663 Kaiserslautern, Germany}
\author{Benito Arnoldi}
\affiliation{University of Kaiserslautern and Research Center OPTIMAS, Erwin-Schr\"odinger-Stra\ss{}e 46, 67663 Kaiserslautern, Germany}
\author{Johannes St\"ockl}
\affiliation{University of Kaiserslautern and Research Center OPTIMAS, Erwin-Schr\"odinger-Stra\ss{}e 46, 67663 Kaiserslautern, Germany}
\author{Florian Haag}
\affiliation{University of Kaiserslautern and Research Center OPTIMAS, Erwin-Schr\"odinger-Stra\ss{}e 46, 67663 Kaiserslautern, Germany}
\author{Ralf Hemm}
\affiliation{University of Kaiserslautern and Research Center OPTIMAS, Erwin-Schr\"odinger-Stra\ss{}e 46, 67663 Kaiserslautern, Germany}
\author{Mirko Cinchetti}
\affiliation{Experimentelle Physik VI, Technische Universit\"at Dortmund, 44221 Dortmund, Germany}
\author{Stefan Mathias}
\affiliation{I. Physikalisches Institut, Georg-August-Universit\"at G\"ottingen, Friedrich-Hund-Platz 1, 37077 G\"ottingen, Germany}
\affiliation{International Center for Advanced Studies of Energy Conversion (ICASEC), Georg-August-Universit\"at G\"ottingen, 37077 G\"ottingen, Germany}
\author{Benjamin Stadtm\"uller}
\affiliation{University of Kaiserslautern and Research Center OPTIMAS, Erwin-Schr\"odinger-Stra\ss{}e 46, 67663 Kaiserslautern, Germany}
\author{Martin Aeschlimann}
\affiliation{University of Kaiserslautern and Research Center OPTIMAS, Erwin-Schr\"odinger-Stra\ss{}e 46, 67663 Kaiserslautern, Germany}



\begin{abstract}
  The high flexibility of organic molecules offers great potential for designing the optical properties of light-active materials for the next generation of optoelectronic and photonic applications.
  However, despite successful implementations of molecular materials in todays’ display and photovoltaic technology, many fundamental aspects of the light-to-charge conversion have still to be uncovered.
  Here, we focus on the ultrafast dynamics of optically excited excitons in C$_{60}$ thin films depending on the molecular coverage and the light-polarization of the optical excitons. 
  Using time- and momentum-resolved photoemission with fs-XUV radiation, we follow the depopulation dynamics in the excited states while simultaneously monitoring the signatures of the excitonic charge character in the molecular valence states.
  Optical excitation with visible light results in the instantaneous formation of charge-transfer (CT) excitons, which transform stepwise into energetically lower Frenkel-like excitons. While the number and energetic position of energy levels within this cascade process are independent of the molecular coverage and the light polarization of the optical excitation, we find quantitative differences in the depopulation times and the optical excitation efficiency.
  Our comprehensive study reveals the crucial role of CT excitons for the excited state dynamics of homo-molecular fullerene materials and thin films.

\end{abstract}

\maketitle
\section{Introduction}
The growing interest in aromatic molecular materials for application-oriented research as well as for fundamental studies is rooted in their exceptional optical properties. Most importantly, the optical band gap of molecular materials can be actively controlled by tuning the molecular structure and composition using chemical synthesis \cite{Henson2012,Hiramoto2014,Schwarze2016}. This allows one to design and optimize the light absorption spectrum of molecular materials for light-harvesting or photovoltaic applications. Despite this prospective of molecular materials for applications, several fundamental aspects of molecular materials are still unresolved, which severely limits our ability to further optimize the functional molecular materials for the next generation of optoelectronic devices.

One of these challenges concerns the light-to-charge conversion process in molecular materials. In contrast to inorganic semiconductors, optical excitation of molecular thin films with visible light does not directly result in the formation of free charge carriers (electrons and holes), but in bound electron-hole pairs called excitons.
These excitons can exhibit different degrees of localization and different spatial charge distributions depending on their excited state energy \cite{Kazaoui1998,Hahn2016,Causa2018}.
For optical excitation energies in the range of the fundamental band gap, the electron-hole pairs are typically located on a single molecular site, i.e., they can be described as Frenkel-like excitons. For larger optical excitation energies, the created excitons can exhibit an at least partial charge-transfer (CT) excitonic character, with electron and hole being separated on neighboring molecular sites. These CT excitons are particularly interesting, since they can act as precursor states for the charge separation and the generation of free charge carriers from excitons.

So far, the formation and decay processes of CT excitons are investigated most frequently at heteromolecular interfaces between donor and acceptor molecules \cite{Zhu2009,Chan2011,Jailaubekov2013,Bernardo2014,Vandewal2014,Abramavicius2016}.
In contrast, only a few studies so far focused on the role of CT excitons for the decay cascade of excitons within homo-molecular structures and thin films, despite their crucial role for the exciton decay process \cite{Hahn2016,Matheson2019,Causa2018,Stadtmuller2019}. This is mainly because the spectroscopic signatures of Frenkel-like and CT-like excitons are extremely difficult to distinguish experimentally.

In this context, we recently uncovered a new and clear signature of CT excitons formed in molecular solids using time-resolved photoemission experiments with fs radiation in the extreme ultraviolet (XUV) spectral range \cite{Stadtmuller2019}. In such an experiment, the large photon energy of the fs-XUV radiation ($22.2 \unit{eV}$ in our case) allows one to simultaneously monitor the exciton dynamics in the excited states and the transient band structure dynamics of the (occupied) molecular valence band structure of the molecules surrounding the optically generated exciton. For a thin film of the prototypical molecule C$_{60}$, we observed the formation and the subsequent population decay cascade of such optically generated excitons in the excited states. These exciton dynamics coincide with a transient spectral broadening of all molecular features in the valence band,  following the population dynamics of the exciton decay process. In our theoretical model simulations, this transient broadening of all valence states could only be explained by the existence of a microscopic transient charge defect in the molecular films, i.e., the transient broadening of all valence states can be identified as spectroscopic signature of CT excitons in molecular films.

Taking advantage of this signature of CT excitons, we were able to identify the spatial charge distribution of the optically excited excitons in a C$_{60}$ thin film. Resonant optical excitation with $3.2 \unit{eV}$ photons instantaneously results in the formation of CT excitons with the hole located in the highest occupied molecular orbital (HOMO) on one molecular site and the electron in the second lowest unoccupied molecular orbital (LUMO+1) on a neighboring molecular site. This CT exciton directly decays into a lower lying CT exciton before transforming into a Frenkel-like exciton with electron and hole on identical molecular sites. Our findings are in agreement with a complementary investigation by Causa \textit{et al.}, reporting similar time scales of the formation and relaxation processes of CT excitons in C$_{60}$ thin films \cite{Causa2018}.

In this work, our recent study of the dynamics of CT and Frenkel excitons in thin C$_{60}$ films on Ag(111) has been extended. Using the same experimental approach as introduced in Ref.~\cite{Stadtmuller2019}, the ultrafast dynamics of CT and Frenkel excitons in thin C$_{60}$ films have been explored for various film thicknesses and light polarizations. Finally, it has been demonstrated that the transient broadening can be observed in the complete valence band structure of the C$_{60}$ thin films throughout the entire Brillouin zone.
Our results will further underline the important role of CT excitons for optical properties of molecular thin films.

\section{Experimental Details}

\subsection{Sample Preparation}
All sample preparation steps have been performed under ultrahigh vacuum (UHV) conditions. The Ag(111) single crystal substrate has been cleaned by several cycles of Ar$^+$-sputtering and subsequent sample annealing at $730 \unit{K}$.
The cleanness of the sample surface has been confirmed by sharp diffraction maxima in low energy electron diffraction (LEED) experiments and by monitoring the existence and linewidth of the Shockley surface state at the $\bar{\Gamma}$-point of the surface Brillouin zone.
The C$_{60}$ molecules have been evaporated onto the clean Ag(111) surface at a pressure $<10^{-9} \unit{mbar}$ using a Knudsen-type evaporation source. The molecular flux used during the evaporation process was calibrated using a quartz crystal oscillator gauge.
The gauge has been calibrated before the experiment by monitoring the peak positions of the C$_{60}$ valence band structure as well as of the work function of the C$_{60}$ thin film and comparing both quantities with values from literature \cite{Weaver1991,Dresselhaus1996}.
The crystalline structure of the C$_{60}$ thin film has been checked with LEED \cite{Li2009}.

\subsection{Time- and Angle-Resolved Photoelectron Spectroscopy}
For the time- (and angle-) resolved photoemission experiments, a hemispherical electron spectrometer (SPECS Phoibos 150), a high-precision six-axis manipulator and a fs-extreme ultraviolet (fs-XUV), $22.2 \unit{eV}$ light source have been combined.

The fs-XUV radiation is obtained by high harmonic generation (HHG). The detailed description of our setup can be found in Ref.~\cite{Eich2014}. In short, our fs-XUV light source is based on a regenerative titanium sapphire chirped-pulse amplifier with sub-$50 \unit{fs}$ pulse duration, $10 \unit{kHz}$ repetition rate and a pulse energy of $1 \unit{mJ}$ at $780 \unit{nm}$ wavelength. Typically, $90$\% of the beam intensity is used for the HHG process. First, the radiation of the laser amplifier systems is frequency-doubled in a $\beta$-barium borate (BBO) crystal and subsequently focused into a hollow waveguide, filled with $30 \unit{Torr}$ of Kr, where the HHG process takes place \cite{Popmintchev2010}.
In our experimental conditions, the high harmonic spectrum exhibits a strong emission line at $22.2 \unit{eV}$ ($7^{\unit{th}}$ harmonic of the HHG spectrum), which is separated by  $6.4 \unit{eV}$ from the next emission lines ($5^{\unit{th}}$ and $9^{\unit{th}}$ harmonic of the HHG spectrum). The HHG radiation is linearly polarized and the orientation of the light polarization (p- or s- polarization with respect to the sample surface) can be adjusted by controlling the light polarization of laser radiation driving the HHG. After guiding the fs-XUV radiation to set of transmissive Al and Sn filters, it is focused onto the sample surface using a toroidal mirror.

The remaining $10\,$\% of the beam intensity of the titanium sapphire amplifier ($\approx 0.1 \unit{mJ}$) is available for the optical excitation of the C$_{60}$ films. This part of the beam is frequency-doubled in a second BBO crystal (photon energy: $3.2 \unit{eV}$) and is focused on the sample surface.
The polarization of the pump pulse is adjusted with a combination of a linear polarizer and a half-wave plate. The delay between the optical pump and the fs-XUV probe pulse is controlled by a delay stage operating with $1 \unit{\mu m}$ longitudinal resolution.

Prior to each experiment, the spatial and temporal overlap of pump and probe pulse on the sample surface was carefully adjusted and checked regularly during the series of measurements. Furthermore, the film has been checked regularly for radiation-induced degradation or dimerization \cite{Lopinski1995}.

In our experiments, we selected analyzer operating parameters which allowed us to record an energy vs. momentum range of $6.4 \unit{eV}$ vs.~$0.8\,${\AA}$^{-1}$ in a single acquisition. Additional momentum-resolved photoemission data were obtained by turning the azimuthal and polar angle of our six-axis manipulator system.

\section{Results and Discussion}
\subsection{Coverage-Dependent Exciton Dynamics}
\begin{figure}[H]
  \includegraphics[width=90mm]{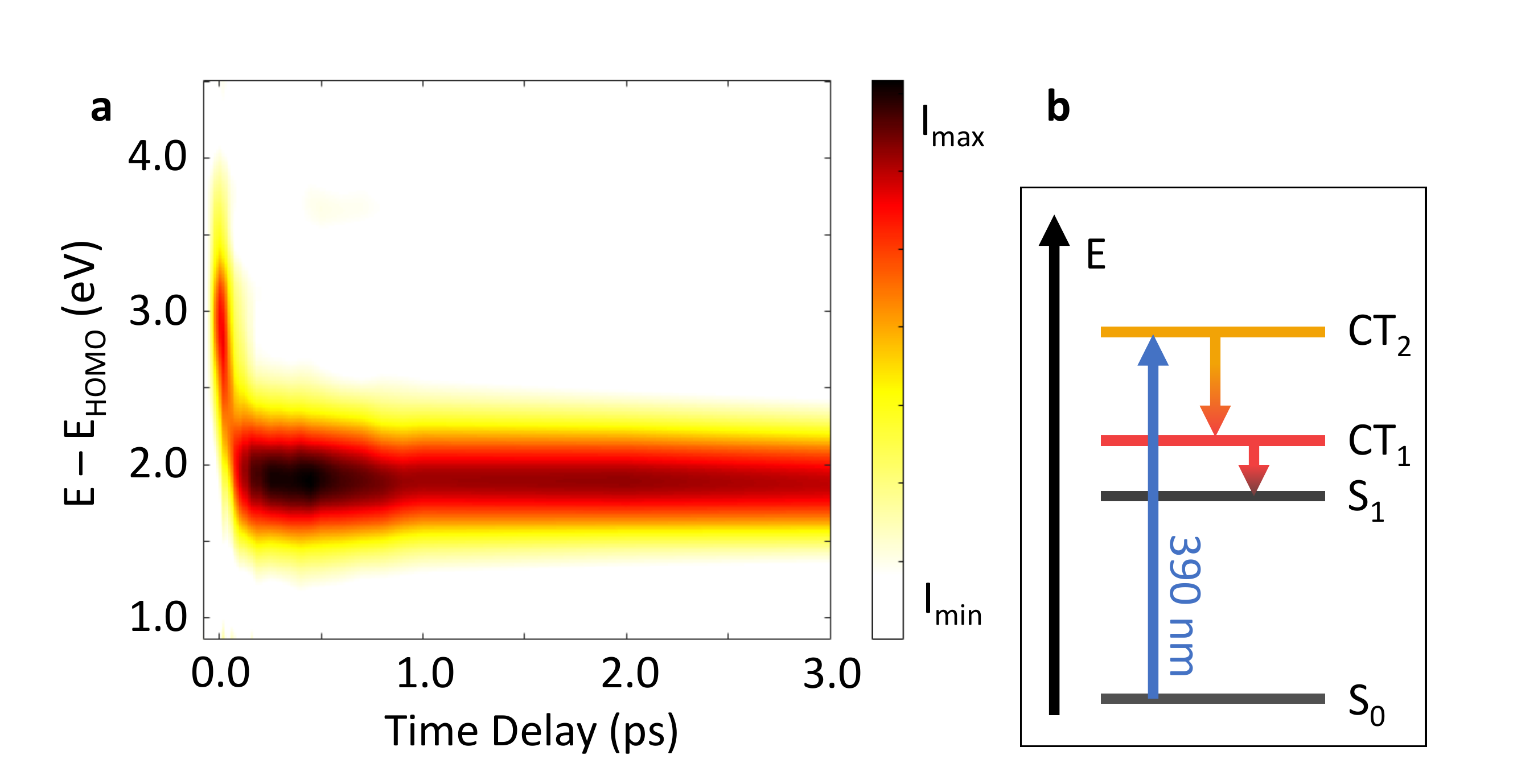}
  \caption{\textbf{a} 2D illustration of the time-resolved photoemission data of the excited state energy range of a C$_{60}$ film after optical excitation with fs-light pulses ($h \nu = 3.2 \unit{eV}$). \textbf{b} Energy level diagram of the excitonic states of the C$_{60}$ thin film.}
  \label{fgr:ov}
\end{figure}

We first focus on the coverage-dependent exciton dynamics of C$_{60}$ thin films on an Ag(111) crystal. The molecular coverage does not only determine the intrinsic electronic properties of the molecular film, but also the coupling between the molecules of the thin film and their environment, for instance, the underlying metallic substrate. The latter is particularly important for energy and momentum dissipation processes and hence ultimately for the excited state dynamics of optically excited excitons in molecular materials \cite{Link2001,Jacquemin1998,Rosenfeldt2010,Dutton2005,Dutton2004,Zhu2004}.

Here, we investigated the excited state dynamics as well as the transient evolution of the molecular valence band structure for molecular coverages between $2.0 \unit{ML}$ and $20 \unit{ML}$. As studied in our previous work \cite{Stadtmuller2019}, the time scales of the dynamics are fluence-independent within the applied range of fluences ($10 - 100 \unit{\mu J\,cm}^{-2}$). We have shown that the fluence only affects the number of excited states, meaning that we are working in the single excitation limit, where influences of higher order effects as trion or bi-exciton formation are neglectable.
For all coverages, we find a qualitatively identical excited state dynamics after optical excitation.
The corresponding exciton dynamics is shown exemplarily in the 2D spectrum of the excited states of a C$_{60}$ film ($\Theta_{\unit{C}_{60}} = 15 \unit{ML}$) and is summarized in the energy level diagram in Fig.~\ref{fgr:ov}. This diagram and the assignment of the excitonic states is based on our recently proposed exciton model for C$_{60}$ thin films in Ref.~\cite{Stadtmuller2019}.
Optical excitation with $3.2 \unit{eV}$ photons leads to an instantaneous formation of an exciton with pronounced charge-transfer character at $E -  E_{\unit{HOMO}} \approx 2.8 \unit{eV}$, labelled CT$_2$. These excitons decay almost instantaneously ($\tau_{\unit{CT}_2} < 100 \unit{fs}$) and repopulate another excitonic level with partial CT character at lower energies (referred to as CT$_1$ exciton).
Finally, the population of the CT$_1$ level decays further into a Frenkel-like exciton state (referred to as S$_1$-exciton) where the excitons are trapped for ns to $\mu$s \cite{Jacquemin1998,Rosenfeldt2010}.

\begin{figure*}
  \includegraphics[width=165mm]{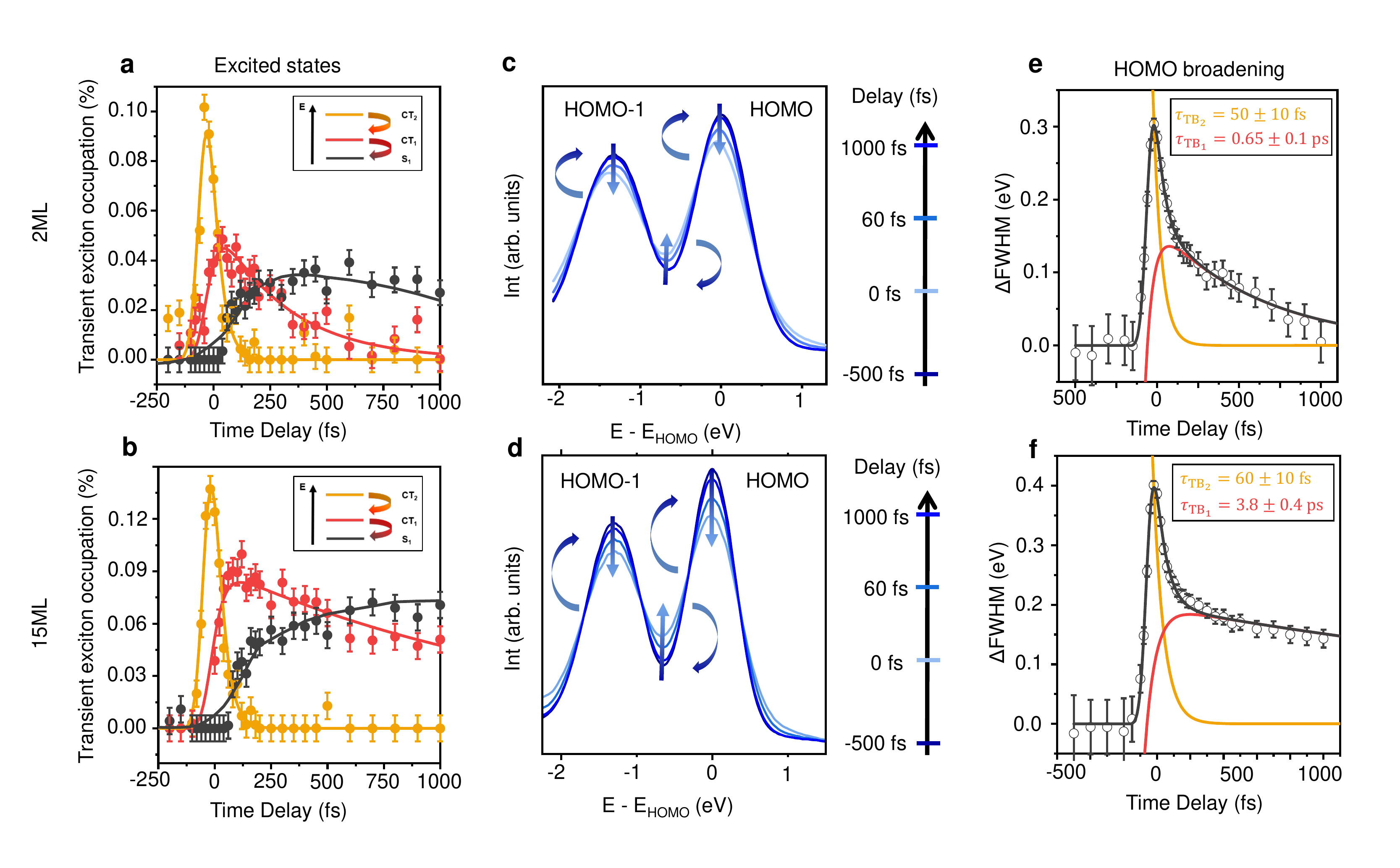}
  \caption{
  \textbf{a,b} Femtosecond transient population dynamics of the excitonic levels of the $2 \unit{ML}$ film (\textbf{a}) and the $15 \unit{ML}$ film (\textbf{b}), showing the extracted dynamics of populations for CT$_2$, CT$_1$ and S$_1$. While the decay time of CT$_2$ stays almost constant, the decay times of CT$_1$ and S$_1$ strongly increase with the C$_{60}$ coverage.
  \textbf{c,d} Energy distribution curves for selected pump-probe time delays, showing the transient broadening of the HOMO levels of the $2 \unit{ML}$ film (\textbf{c}) and the $15 \unit{ML}$ film (\textbf{d}). \textbf{e,f} Transient broadening of the HOMO levels, extracted by a spectral analysis from the energy distribution curves shown in \textbf{c,d} of the $2 \unit{ML}$ film (\textbf{e}) and the $15 \unit{ML}$ film (\textbf{f}).
  Following the dynamics of the excited states in the population dynamics (\textbf{a,b}), the fast decay constant $\tau_{\unit{TB}_2}$ is independent on the C$_{60}$ coverage, while the slow decay constant $\tau_{\unit{TB}_1}$ strongly increases with the film thickness.}
  \label{fgr:thickness1}
\end{figure*}

The population dynamics of the excitons are quantified by analyzing the spectral density of the time-resolved photoemission data set of the excited states using the same fitting procedure as developed in Ref.~\cite{Stadtmuller2019}. The resulting transient population of the CT$_2$-, CT$_1$- and S$_1$-exciton are shown exemplarily in Fig.~\ref{fgr:thickness1}a,b for molecular coverages of $\Theta_{\unit{C}_{60}} = 2.0 \unit{ML}$ and $\Theta_{\unit{C}_{60}} = 15 \unit{ML}$.
The data points including the experimental uncertainty are shown in orange, red and dark red dots for the \mbox{CT$_2$-,} CT$_1$- and S$_1$-exciton.
The corresponding dynamics of each level are fitted by two exponential functions to model its population $\bar{\tau}$ and depopulation $\tau$ time. The best fitting results are included in Fig.~\ref{fgr:thickness1}a,b as solid lines of identical color as the data points.
For both coverages, we find extremely fast depopulation time constants of the CT$_2$ exciton of $\tau_{\unit{CT}_2}(2 \unit{ML}) = 40 \pm 10 \unit{fs}$ and $\tau_{\unit{CT}_2}(15 \unit{ML}) = 38 \pm 10 \unit{fs}$ for the $2 \unit{ML}$ film and $15 \unit{ML}$ film, respectively.
For each coverage, these depopulation time constants $\tau_{\unit{CT}_2}$ are identical to the population time constants $\bar{\tau}_{\unit{CT}_1}$ of the energetically lower CT$_1$ exciton, suggesting a direct refilling from the CT$_2$ to the CT$_1$ exciton.
In contrast to the CT$_2$ exciton, the depopulation times of the CT$_1$ exciton are significantly different for both coverages and reveal values of $\tau_{\unit{CT}_1}(2\unit{ML}) = 310 \pm 70 \unit{fs}$ and $\tau_{\unit{CT}_1}(15 \unit{ML}) = 1.5 \pm 0.3 \unit{ps}$, i.e., the depopulation time $\tau_{\unit{CT}_1}$ increases with increasing coverage.
Similarly, the depopulation time of the Frenkel exciton $\tau_{\unit{S}_1}$ also increases with increasing coverage, in agreement with recent studies of the exciton dynamics of molecular materials \cite{Link2001,Rosenfeldt2010,Shibuta2016}.

To determine the charge distribution of these single quasi-particle dynamics, we need to study their influence the ensemble. Therefore, we now move to the discussion of the transient changes of the whole valence band structure. Energy distribution curves in the energy region of the HOMO and the HOMO-1 state at selected time delays are shown in Fig.~\ref{fgr:thickness1}c,d for both molecular coverages.
In both cases, we observe an instantaneous transient increase of the linewidth of both molecular states upon optical excitation, which reduces again on the picosecond timescale.
This transient broadening is clearly induced by the formation of CT excitons in both C$_{60}$ films upon optical excitation, independent from the molecular coverage. To quantify the dynamics of the CT excitons in the C$_{60}$ film, we analyze the transient broadening of the molecular valence band structure using the fitting procedure established in Ref.~\cite{Stadtmuller2019}.
The extracted transient broadening curves for the HOMO level for both molecular films are shown in Fig.~\ref{fgr:thickness1}e,f. For both cases, the best fitting result is obtained by using a double-exponential decay function with decay constants that are comparable with the depopulation times $\tau_{\unit{CT}_2}$ and $\tau_{\unit{CT}_1}$, obtained for the corresponding molecular film.
In particular, we find $\tau_{\unit{TB}_2}(2 \unit{ML}) = 50 \pm 10 \unit{fs}$, $\tau_{\unit{TB}_1}(2 \unit{ML}) = 650 \pm 100 \unit{fs}$, $\tau_{\unit{TB}_2}(15 \unit{ML}) = 60 \pm 10 \unit{fs}$, and $\tau_{\unit{TB}_1}(15 \unit{ML}) = 3.8 \pm 0.4 \unit{ps}$.
Most importantly, these time scales are significantly shorter than the decay constant of the S$_1$-exciton mentioned earlier, being in the order of ns to $\mu$s. Consequently, the transient broadening is not affected by the dynamics of the S$_1$-exciton. This allows us to conclude that only the excitonic levels CT$_2$ and CT$_1$ reveal a significant CT exciton character, while the S$_1$-exciton exhibits exclusive Frenkel exciton-like character.

\begin{figure}
  \includegraphics[width=70mm]{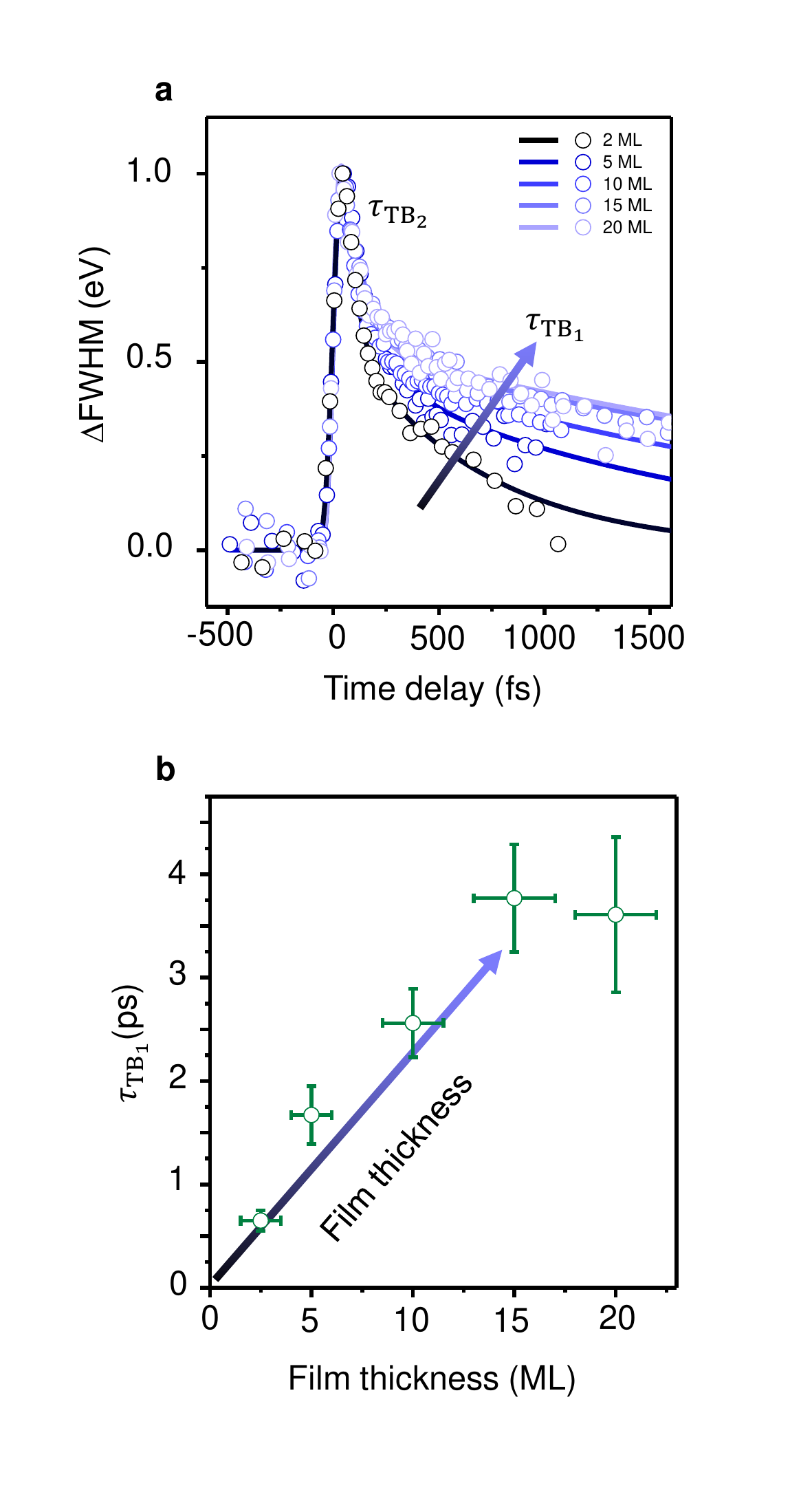}
  \caption{
  \textbf{a} The transient linewidth broadening of the valence states, extracted for several film thicknesses. They show a constant fast decay time $\tau_{\unit{TB}_2}$ and an increase of the slow decay time $\tau_{\unit{TB}_1}$ with increasing film thickness. \textbf{b} Correlation between the slow decay time $\tau_{\unit{TB}_1}$ and the C$_{60}$ coverage. We find a monotonic increase of $\tau_{\unit{TB}_2}$, as observed in \textbf{a}, and a potential saturation for coverages above $15 \unit{ML}$. }
  \label{fgr:thickness2}
\end{figure}

The discussed dynamics can be detected for all molecular coverages between $2.0 \unit{ML}$ and $20 \unit{ML}$. This is clearly visible in Fig.~\ref{fgr:thickness2}a, which shows the transient broadening curves of the HOMO level obtained for five molecular films in this coverage regime.
The extracted decay constant $\tau_{\unit{TB}_1}$, corresponding to the CT$_1$ decay ($\tau_{\unit{CT}_1}$), is shown in Fig.~\ref{fgr:thickness2}b. The decay time of the energetically lowest CT exciton (CT$_1$) into the Frenkel exciton-like state S$_1$ increases continuously with increasing coverage before saturating at around $15 \unit{ML} - 20 \unit{ML}$. The asymptotic values for the exciton decay hence reflect the intrinsic dynamics of CT excitons of the fullerene thin film.

Most importantly, the direct comparison of the excited state dynamics and the transient broadening of the valence states also allows us to estimate coverage-dependent changes of the charge-transfer/Frenkel exciton ratio of the optically excitons in the C$_{60}$ thin films. More precisely, the optically excited states will have an at least \textit{partial} CT character, the other part of the optically excited states will have a Frenkel-like excitonic character \cite{Causa2018}. The influence of the film thickness on this CT/Frenkel exciton ratio can be deduced from our experimental data. It is expected to be equivalent to the ratio between the population of the CT$_2$ excitons $P_{\unit{CT}_2} (0 \unit{fs})$ and the transient broadening $\Delta \unit{FWHM}(0 \unit{fs})$ at the instant of the optical excitation ($\Delta t = 0 \unit{fs}$):
\begin{equation}
  R(\Theta_{\unit{C}_{60}}) = \frac{P_{\unit{CT}_2}(0 \unit{fs},\Theta_{\unit{C}_{60}})}{\Delta \unit{FWHM}(0 \unit{fs},\Theta_{\unit{C}_{60}})}.
\end{equation}
The population $P_{\unit{CT}_2}(0 \unit{fs})$ is obtained by the normalized photoemission intensity of the CT$_2$ signal at $\Delta t = 0 \unit{fs}$ and is proportional to the number of optically excited excitons with both CT and Frenkel-like character.
The transient broadening $\Delta \unit{FWHM}(0 \unit{fs})$, on the other hand, is only a measure for the part of optically excited excitons with CT character. Interestingly, we find a constant ratio $R = 0.33 \pm 0.03 \frac{1}{eV}$ for all molecular coverages.
This suggests a coverage-independent ratio between the CT and Frenkel-like character of the optically excited excitons. This finding might not be surprising when considering the band structure of C$_{60}$ thin films. They exhibit an almost atomic crystal-like band structure with strongly dispersing valence states \cite{Latzke2019} that can be identified as one of the important ingredients for the formation of excitons with mixed CT and Frenkel exciton character \cite{Kazaoui1998}.
This type of band structure and energy level alignment is almost identical for molecular coverages beyond $1.0 \unit{ML}$ and only changes significantly for molecules in direct contact with the Ag(111) surface. We hence propose that these only marginal changes in the energy alignment of the molecular valence states and the optical band gap for molecules in the second and higher C$_{60}$ layers are the reason for the observed coverage-independent ratio of the CT and Frenkel exciton-like character of the optically excited CT$_2$ excitons in C$_{60}$ thin films.

In this regard, our investigation of the coverage-dependent exciton dynamics in C$_{60}$ thin films allows us to draw the following conclusions:
\begin{enumerate}
\item{The CT$_2$ and CT$_1$ excitonic states are of at least partial charge-transfer exciton character for the observed range of molecular coverages, while the S$_1$-exciton is of Frenkel-like character.}
\item{The Frenkel-CT exciton ratio of the optically excited CT$_2$ excitons is independent on the molecular coverage.}
\item{The depopulation time $\tau_{\unit{CT}_1}$ of the CT$_1$ exciton increases with increasing molecular coverage.}
\end{enumerate}

Our findings hence do not only reflect the coverage-dependent exciton dynamics of fullerene thin films \cite{Link2001,Rosenfeldt2010,Shibuta2016,Dutton2004,Jacquemin1998} and other molecular complexes \cite{Varene2012} on various substrates. They also shine new light onto the coverage-dependent charge distribution of the optically excited excitons and their coverage-dependent relaxation processes back to the ground state.

Note that minor quantitative differences between our findings and literature can be attributed to the significantly larger probe photon energies used in our experiment compared to typical time-resolved photoemission studies of the exciton dynamics in molecular materials. This large photon energy of $h \nu = 22.2 \unit{eV}$ results in an extremely high surface sensitivity of our experiment, which is typically not the case for low energy two-photon photoemission studies using laser light sources in the visible to ultraviolet range \cite{Dutton2005,Zhu2004}.

\subsection{Light Polarization-Dependent Exciton Dynamics}
\begin{figure*}
  \includegraphics[width=165mm]{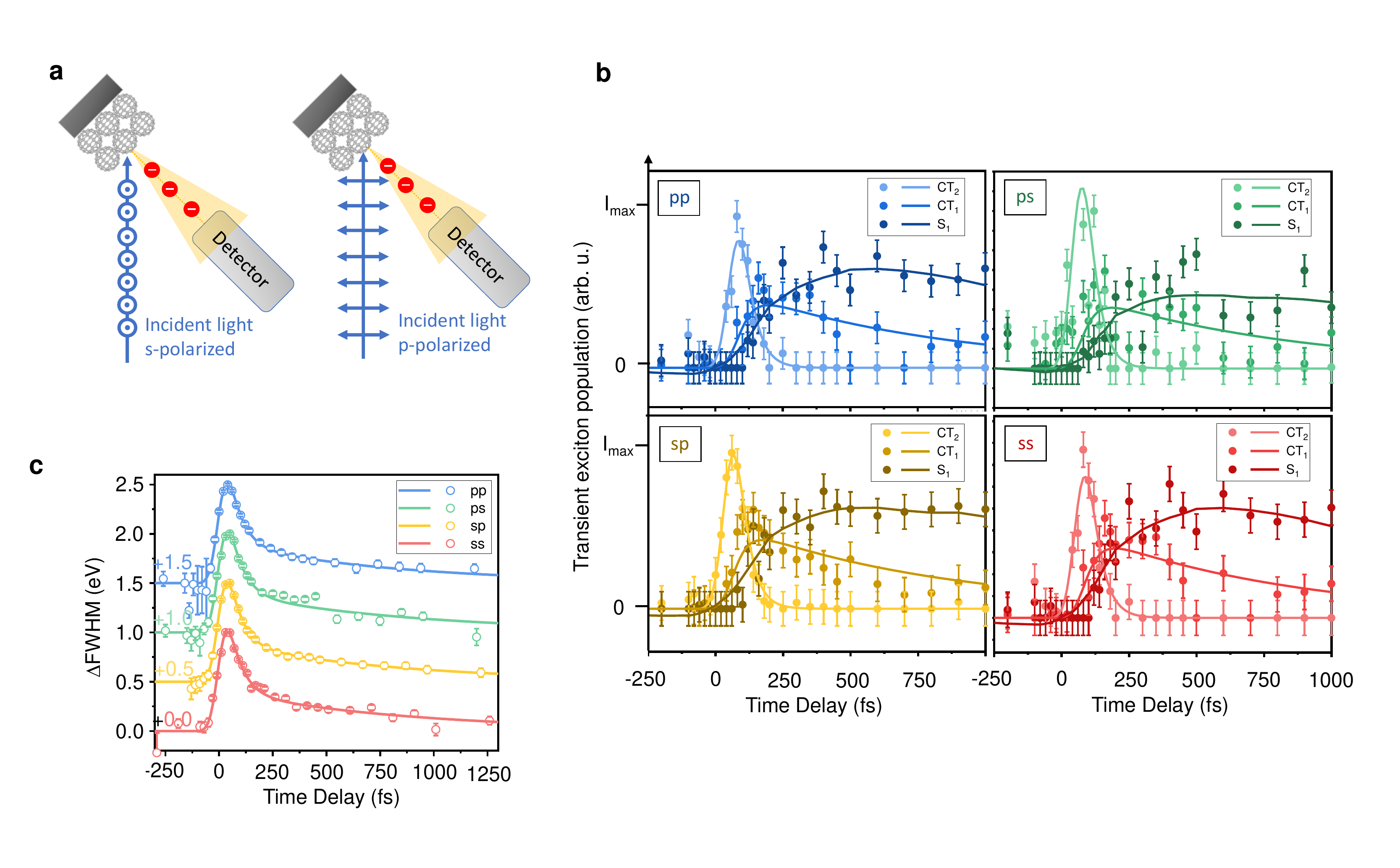}
  \caption{
  \textbf{a} Experimental geometry of the tr-photoemission study for different light polarizations. For s-polarized light (case \textit{left}), the electric field relative to the sample is completely in-plane. For p-polarized light (case \textit{right}), the electric field consists of a mixture of in-plane and out-of-plane components. \textbf{b} Light-induced population dynamics for the different combinations of pump and probe light polarization. The color of the CT$_2$ exciton dynamics reflects the color of the transient broadening in \textbf{c}. \textbf{c} Transient broadening of the HOMO level for different combinations of probe and pump light polarization. The plots have been vertically shifted by $0.5 \unit{eV}$ with respect to each other for better visibility. Similar to the invariance of the population dynamics, also the transient broadening dynamics are not affected by the pump and probe light polarization.  }
  \label{fgr:pol}
\end{figure*}

Next, we turn to the polarization-dependent exciton dynamics of C$_{60}$ thin films. The highly crystalline order of the C$_{60}$ solid results in the formation of a complex molecular band structure with valence bands of different orbital character, despite the spherical symmetry of the C$_{60}$ molecule itself. For instance, the highly dispersing valence states of a C$_{60}$ thin film exhibit hole-like and electron-like dispersions with p$_{x/y}$ and p$_{z}$-like orbital character \cite{Haag2020,Latzke2019}. This variety of orbital characters of the valence states can lead to transitions into different optically excited states depending on the light polarization. As a result, the efficiency of the light absorption and the optically induced exciton formation could be severely altered for different light polarizations. Even more, light polarization-dependent optical transitions could result in the formation of excitons with different CT or Frenkel character.

To explore these aspects, we recorded the excited state dynamics and the transient evolution of the valence states of a C$_{60}$ thin film with a coverage of $\Theta_{\unit{C}_{60}} = 5 \pm 2 \unit{ML}$ for all four possible combinations of light polarizations of pump and probe radiation. Accordingly, the experiment has been repeated for p-polarized probe and p-polarized pump beam (pp), for p-polarized probe and s-polarized pump beam (ps), for s-polarized probe and p-polarized pump beam (sp) and for s-polarized probe and s-polarized pump beam (ss), respectively. In our experiment, we use an incidence angle of $45^{\circ}$ for the pump and probe radiation in normal emission geometry as shown in Fig.~\ref{fgr:pol}a. In this geometry, p-polarized light exhibits an in-plane as well as an out-of-plane component of the electric field vector with respect to the surface plane, while the electric field vector of s-polarized light is located completely in the sample surface plane.

We start the discussion with the light absorption behavior of the C$_{60}$ films for different incident light polarizations of pump and probe. Hereby, the polarization of the probe beam only influences the overall count rate of our experiment, i.e., the photoemission cross section of the C$_{60}$ film is significantly larger for p-polarized light compared to s-polarized light. In contrast, changing the polarization of the optical excitation pulse results in a severe change in the number of optically excited excitons.

Using a photon energy of $3.2 \unit{eV}$, we observe an optically generated population of the CT$_2$ exciton of $P_{\unit{CT}_2, \unit{p-pol}}(0 \unit{fs}) = (3.8 \pm 0.4) 10^{-3}$ for p-polarized light at a laser fluence of $F_{\unit{p}} = 50 \pm 10 \unit{\mu J \, cm^{-2}}$.
For s-polarized light, a significantly larger laser fluence of $F_{\unit{s}} = 70 \pm 10 \unit{\mu J \, cm^{-2}}$ is required to obtain a comparable exciton population of $P_{\unit{CT}_2, \unit{s-pol}}(0 \unit{fs}) = (4.0 \pm 0.4) 10^{-3}$.
The similar exciton population for these significantly different pump fluencies is also confirmed by the almost identical transient linewidth broadening of $\Delta \unit{FWHM}_{\unit{p-pol}}(0 \unit{fs}) = 0.37 \pm 0.03 \unit{eV}$, and $\Delta \unit{FWHM}_{\unit{s-pol}}(0 \unit{fs}) = 0.35 \pm 0.03 \unit{eV}$ for optical excitation with p- and s-polarized light.
Based on these values, we can estimate that the exciton exciting efficiency of s-polarized light is only $61 \pm 5\,\%$ of the exciton excitation efficiency of p-polarized light. This large difference cannot be fully explained by the different absorption efficiency of p- and s-polarized light.
\begin{figure*}
  \includegraphics[width=165mm]{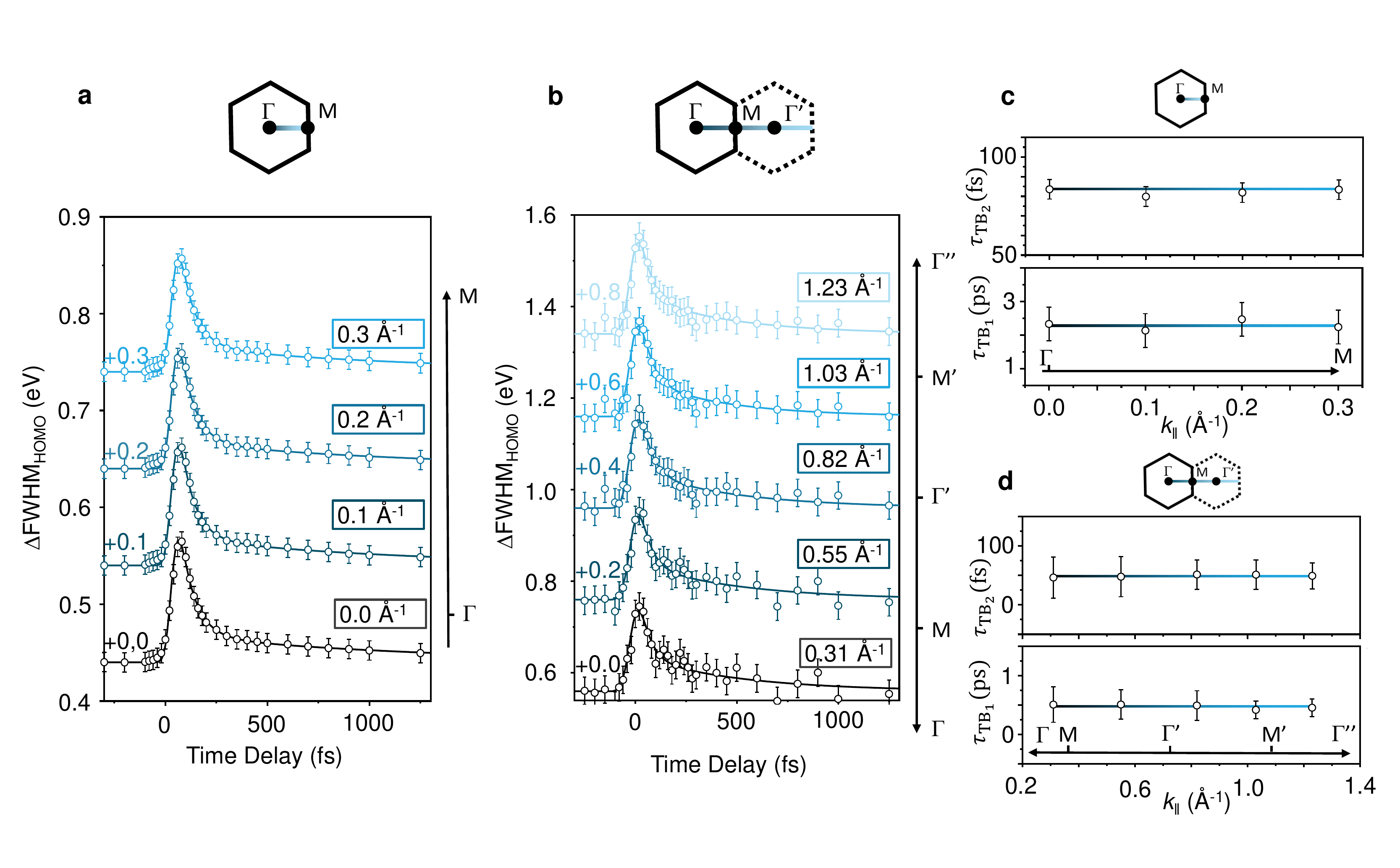}
  \caption{
  \textbf{a} Direct comparison of the transient linewidth broadening of the HOMO width for different momenta within the first Brillouin zone along the $\bar{\Gamma}\bar{\unit{M}}$-direction (see sketch of Brillouin zone above). Each curve has been shifted vertically by $0.1 \unit{eV}$ for better visibility. \textbf{b} Direct comparison of the transient HOMO width for different momenta in $\bar{\Gamma}\bar{\unit{M}}$-direction to the $\bar{\unit{M}}$ point and beyond (see sketch above graph for orientation). The plots have been shifted by $0.2 \unit{eV}$ vertically for better visibility.
  \textbf{c} Extracted decay times $\tau_{\unit{TB}_2}$ and $\tau_{\unit{TB}_1}$ for the transient HOMO width shown in \textbf{a}.
  There is no change within error bars. \textbf{d} Extracted decay times $\tau_{\unit{TB}_2}$ and $\tau_{\unit{TB}_1}$ for the transient HOMO width shown in \textbf{b}. There is no change within error bars.
  Note that the measurements for (\textbf{a,c}) and (\textbf{b,d}) were conducted on different sample preparations. Thus, the qualitative change of decay times between both measurements has its origin in different C$_{60}$ coverages between the two C$_{60}$ films as has been discussed previously.}
  \label{fgr:Bri}
\end{figure*}

The relative absorption of p- and s-polarized light can be calculated from the index of refraction $n(\hbar \omega) = n_r(\hbar \omega) + i n_i(\hbar \omega)$, with the values $n_r(3.2 \unit{eV}) = 2.5 \pm  0.1$ and $n_i(3.2 \unit{eV}) = 1.25 \pm 0.20$ taken from literature \cite{Yagi2009}.
Using the Fresnel equations, we find that the adsorption of s-polarized light in a C$_{60}$ thin film is only $10 \pm 5\,\%$ smaller compared to p-polarized light. This prediction is significantly smaller compared our result and hence suggests that the difference in the optical excitation efficiency of CT excitons in C$_{60}$ thin films cannot only be due to the different light adsorption of both polarizations. Instead, other effects such as the symmetry of the valence and excited state wave function or their orbital character must also contribute to the optical excitation efficiency of excitons in C$_{60}$ thin films. As mentioned earlier, the excitation fluence does not influence the time scales of the exciton dynamics \cite{Stadtmuller2019}. Therefore, the differences in light absorption efficiency can be compensated by adapting the excitation fluence without influencing the exciton dynamics.

Having discussed the light absorption behavior of the C$_{60}$ thin film, we will now present the femtosecond dynamics of the sample for the different combinations of pump and probe polarization.
For all four combinations of light polarization, we extracted the population dynamics of the CT$_2$, CT$_1$ and S$_1$ excitonic state using the same fitting routine as discussed above. The transient populations of these states are shown in Fig.~\ref{fgr:pol}b.
The population and depopulation time constants $\bar{\tau}$ and $\tau$ are obtained individually for each state using exponential functions. The best fitting results are included in Fig.~\ref{fgr:pol}b as solid lines.

The optical excitation results in an instantaneous formation of the CT$_2$ exciton. Subsequently, it transforms extremely fast into the CT$_1$ excitons with $\tau_{\unit{CT}_2} = \bar{\tau}_{\unit{CT}_1} = 45 \pm 10 \unit{fs}$. The decay time of the CT$_1$ state is $\tau_{\unit{CT}_1} = 0.8 \pm 0.3 \unit{ps}$ which is consistent with our coverage-dependent study of the exciton dynamics of C$_{60}$ thin films. Note that the population and depopulation decay constants are identical for all four experiments within the experimental uncertainty. This points to polarization-independent population dynamics of the excitons in C$_{60}$.

A similar behavior can be observed for the transient linewidth broadening of the molecular valence states. In all four cases, the optical excitation instantaneously results in a transient linewidth broadening which is quantified using a Gaussian fitting model (details see above or in Ref.~\cite{Stadtmuller2019}). The evolution of the transient broadening $\Delta \unit{FWHM}$ of the HOMO level is shown for all four combinations of polarization in Fig.~\ref{fgr:pol}c, revealing a double-exponential decay with identical decay constants of $\tau_{\unit{TB}_2}= 60 \pm 10 \unit{fs}$ and $\tau_{\unit{TB}_1} = 0.8 \pm 0.2 \unit{ps}$.

Therefore, we can conclude that neither the spatial charge distribution nor the time scales of the excited excitons depend on the polarization of the optically excited light pulses. Moreover, it is interesting to point out that the transient linewidth broadening is also independent of the polarization of the probe radiation. This shows that the excitons of mixed CT- and Frenkel-character equally affect molecular valence states of different orbital character, i.e, of p$_{x/y}$-character observed with s-polarized light and of p$_{z}$-character observed predominantly with p-polarized light.
\\
Interestingly, the light polarization-independent depopulation dynamics of the CT excitons reported above allow us to further investigate the transient linewidth broadening of the valence states throughout the surface Brillouin zone of the C$_{60}$ crystal structure.
In a typical time- and angle-resolved photoemission experiment with linearly polarized light and a fixed angle between the incident light and the photoemission detector, the orientation of the electrical field vector of the pump and probe radiation changes continuously when changing the emission angle of the photoelectrons.
This results in a continuous modification of the optical excitation conditions (pump pulse) or the corresponding photoemission cross section (probe pulse), making it extremely difficult to quantify the experimental observables.
This challenge can only be prevented when aligning the electric field vector of the pump and probe pulse with the rotation axis of the sample surface, i.e., by using s-polarization for the pump and probe pulse.
For most experiments and material systems, this severely limits the accessible information. In our case, the light-polarization independent dynamics allow us to fully access the complete information regarding the exciton dynamics in momentum space using only s-polarized light both for the optical excitation and for the photoemission of excited as well as the molecular valence states.

Taking advantage of this opportunity, we recorded time- and momentum-resolved photoemission data within in the first surface Brillouin zone along the $\bar{\Gamma}\bar{\unit{M}}$-direction of the C$_{60}$ crystal structure. At room temperature, C$_{60}$ forms a long range ordered $(2 \sqrt{3} \times 2 \sqrt{3}) 30^{\circ}$ superstructure with two additional structural domains rotated by $\pm 15^\circ$ \cite{Li2009}.
The corresponding surface Brillouin zone of the central C$_{60}$ domain is shown in Fig.~\ref{fgr:Bri}a, the $\bar{\Gamma}\bar{\unit{M}}$-direction is marked by a blue line. Time-resolved photoemission data sets were recorded in steps of $0.1\,$\AA$^{-1}$ and the transient line shape of the entire valence band structure was analyzed using the same fitting model as discussed before.

The transient linewidth broadening of the HOMO level is shown in Fig.~\ref{fgr:Bri}a. For all points in the surface Brillouin zone, we observe an identical temporal evolution of the HOMO linewidth, similar to our findings discussed above. The HOMO linewidth increases instantaneously upon optical excitation, followed by a double-exponential decay. The decay constants of the double-exponential decay are $\tau_{\unit{TB}_2}= 82 \pm 10 \unit{fs}$ for the first decay step and $\tau_{\unit{TB}_1}= 2.3 \pm 0.5 \unit{ps}$ for the second one.
Interestingly, the magnitude of the transient broadening, i.e., the maximum transient linewidth broadening at $\Delta t = 0 \unit{fs}$, $\Delta \unit{FWHM}(0 \unit{fs})$, is also identical for all momenta along the $\bar{\Gamma}\bar{\unit{M}}$-direction.

For a second C$_{60}$ sample, we repeated the same experiment in a large range of momenta covering also the second and third surface Brillouin zone of the C$_{60}$ crystal. The transient linewidth broadening traces of the HOMO are shown in Fig.~\ref{fgr:Bri}b. All traces exhibit an identical shape, i.e., an instantaneous rise with an identical maximal linewidth broadening $\Delta \unit{FWHM}(0 \unit{fs})$, followed by a double-exponential decay with identical decay constants of $\tau_{\unit{TB}_2}= 50 \pm 25 \unit{fs}$ and $\tau_{\unit{TB}_1} = 0.50 \pm 0.25 \unit{ps}$.

The momentum-dependent decay times of the transient linewidth broadening of the C$_{60}$ HOMO are summarized in Fig \ref{fgr:Bri}c,d.
They clearly show that the transient linewidth broadening occurs simultaneously throughout the entire valence band structure with an equal amplitude $\Delta \unit{FWHM}(0 \unit{fs})$. The transient broadening itself thus only depends on the film thickness (as discussed earlier) and the strength of the optical excitation (discussed in \cite{Stadtmuller2019}). These observations are fully in line with our proposed model for the transient broadening \cite{Stadtmuller2019}: The optically excited CT excitons create a transient electrostatic (Stark-like) field that results in a transient polarization of the surrounding molecular materials.
In turn, this transient change in the dielectric screening results in a transient energy shift of the molecular valence states of the organic molecules surrounding the CT exciton. Crucially, the magnitude and sign of the transient band structure renormalization only depend on the relative position of the CT exciton and the probed molecule.

In particular, our model predicts an equal shift of all molecular states, independent of their momentum. In this regard, our experiments are fully supported by our proposed model.

\section{Conclusion}
In this work, we investigated the dynamics of CT and Frenkel excitons in thin C$_{60}$ films on Ag(111), depending on the molecular coverage as well as of the light polarization of the optical exciton. Using time- and momentum-resolved photoemission with fs-XUV radiation, we followed the population dynamics of the excitons in the excited states, while simultaneously monitoring the transient signatures of the charge character of the excitons in the molecular valence states.
We showed that the optical excitation of C$_{60}$ thin films results in the direct formation of an exciton with dominant CT character for all molecular coverages between $2.0 \unit{ML}$ and $20 \unit{ML}$. This CT exciton (CT$_2$) subsequently decays stepwise into an energetically lower CT exciton (CT$_1$) on sub-$100 \unit{fs}$ time scales for all coverages before transforming into a Frenkel-like exciton.
The depopulation time of the CT$_1$ exciton increases with increasing coverage and saturates at $\tau_{\unit{CT}_1} \approx 3.5 \unit{ps}$ for molecular coverages above $15 \unit{ML}$.
Crucially, the CT-Frenkel exciton ratio of the optically excited CT$_2$ exciton does not depend on the molecular coverage.
Similarly, we did not observe any modification of the exciton population decay dynamics under change of the optical excitation from p- to s- polarized light. The latter only results in a reduced number of optically excited excitons due to a reduction of the optical excitation efficiency of excitons in C$_{60}$ films.

In this way, our comprehensive study of the exciton dynamics of fullerene thin films provides a clear view onto transient population decay and the charge character of excitons in molecular thin films. In particular, we demonstrate the crucial role of CT excitons even for the excited state dynamics of homo-molecular fullerene materials and thin films.

\begin{acknowledgments}
  The research leading to these results was funded by the Deutsche Forschungsgemeinschaft (DFG, German Research Foundation) - TRR 173 - 268565370, project B05. B.S. and S.E. also acknowledge financial support from the Graduate School of Excellence Mainz (Excellence initiative DFG/GSC 266). This work is supported by the European Research Council (Grant 725767-hyControl).
\end{acknowledgments}

\providecommand{\noopsort}[1]{}\providecommand{\singleletter}[1]{#1}%


\end{document}